\documentclass[aps,prb,twocolumn,superscriptaddress]{revtex4}
\usepackage{graphicx}

\begin{document}

\title{Stability of the shell structure in 2D quantum dots}
\author{M.~Aichinger}
\email[Electronic address: ]{aichingerm@gmx.at}
\affiliation{Institut f\"ur Theoretische Physik, Johannes Kepler 
Universit\"at, A-4040 Linz, Austria}
\author{E.~R\"as\"anen}
\email[Electronic address: ]{ehr@fyslab.hut.fi}
\affiliation{Laboratory of Physics, Helsinki University of Technology,
P.O. Box 1100, FIN-02015 HUT, Finland}

\date{\today}

\begin{abstract}
We study the effects of external impurities on the 
shell structure in semiconductor quantum dots by 
using a fast response-function
method for solving the Kohn-Sham equations. We perform 
statistics of the addition energies up to 20 interacting electrons.
The results show that the shell structure is generally
preserved even if effects of high disorder are clear.
The Coulomb interaction and the 
variation in ground-state spins have a strong effect on
the addition-energy distributions, which in the
noninteracting single-electron picture correspond to level 
statistics showing mixtures of Poisson and Wigner forms. 
\end{abstract}

\maketitle

\section{Introduction}

The functionality of nanoelectronic devices strongly depends on
the quality of the samples produced in the fabrication process.
In semiconductor quantum dots (QD),~\cite{qd} there may exist 
external impurities, which have considerable effects on the 
measurable quantities such as the addition energy and the chemical 
potential. Generally, the physical properties of QD's can be determined
by the shape of the confining potential restricting the electrons in the
semiconducting material. 

In a vertical QD formed in a double-barrier 
heterostructure, the measured addition energies
(cf. electron affinities) correspond well to the shell
structure of a two-dimensional parabolic (harmonic) model 
potential.~\cite{tarucha} Remarkably good agreement between the 
experiment and theory has been obtained by using this model and 
taking the electron-electron interactions into account by 
employing, e.g., the spin-density-functional theory 
(SDFT).~\cite{hirose2,reimann} 
There are, however, always sample-specific variations~\cite{matagne}
such that in QD's with more than only a few electrons, 
the agreement between the experiments and theory 
becomes generally worse, presumably due to non-symmetric 
deviations of the actual confining potential.

Deviations from the circular symmetry can be induced by 
impurity particles migrated through the heterostructure layers. 
States bound to hydrogenic impurities were found in the single-electron
tunneling experiment by Ashoori {\em et al.}~\cite{ashoori}
Recently, a clear deformation in the transport spectrum of a
resonant tunneling device was shown to be due to an ionized
single or double-acceptor near the QD.~\cite{esa6}
For the time being, there is no experimental data showing systematic 
impurity effects, but the eventual
serial production of QD structures can lead to a considerable
fraction of distorted samples. 

Theoretical research of QD's containing impurities has been
mostly based on either single-impurity systems studied with exact 
diagonalization~\cite{halonen} and the SDFT,~\cite{esa6} or
on QD's with a relatively high impurity 
density.~\cite{hirose1,gycly} 
In the latter case, Hirose {\em et al.}~\cite{hirose1} have
performed SDFT studies on the energy fluctuations 
and the effects of interactions, as well as on the role of 
spin.~\cite{hirose3} 
These systems have experimental realization
in disordered (no shell structure) QD's, where the 
fluctuations in the Coulomb blockade peak spacings show 
interesting statistics beyond the constant interaction model 
combined with the random matrix theory.~\cite{sivan,berkovits}

In this paper we focus on the stability of the shell structure
in primarily parabolic few-electron QD's. Particularly,
the effects of external impurities on the addition energies 
are studied up to the high-disorder limit. The two main
extensions to previous studies are (i) the use of an impurity model 
based on recent experimental findings~\cite{esa6} and (ii) the 
computation of the complete addition energy spectrum up to 20 
electrons with both a single impurity at an arbitrary position, 
and finally with an ensemble of one thousand impurity 
configurations. The calculations are performed using the SDFT 
with a remarkably fast response-function iteration method for 
the Kohn-Sham (KS) equations.~\cite{michael} The resulting spectra
demonstrate that the shell structure is generally preserved,
although in the mid-shell regime there can be drastic changes in the 
energetics. The statistics of the 
noninteracting level spacings can be characterized by a mixture of 
Poisson and Wigner distributions. Those shapes are strongly affected 
by the Coulomb interaction and by the ground-state spins.

The organization of this paper is as follows. First in Sec.~\ref{model} 
we define the model
system and the parameters for the external impurities. In Sec.~\ref{method}
we introduce the response-function scheme for the density-functional
calculations. The results for a single repulsive impurity are presented
in Sec.~\ref{results1}. Finally the addition-energy distributions from
one thousand configurations are shown in Sec.~\ref{results2}.
The paper is summarized in Sec.~\ref{summary}.

\section{Model} \label{model}

We model the quantum dot by using the conventional effective-mass
approximation and choose the material parameters for GaAs,
i.e., the effective mass $m^*=0.067\,m_e$ and the dielectric
constant $\epsilon=12.7$. We apply a strictly 
two-dimensional model by assuming a negligible degree
of freedom for electrons in the vertical direction.
The many-electron Hamiltonian is given in the effective atomic 
units~\cite{units} as
\begin{equation}
H=\sum^N_{i=1}\left[-\frac{\nabla^2_i}{2}+
\frac{1}{2}\omega_0^2 r_i^2+V_{\rm imp}({\mathbf r}_i)\right]+
\sum^N_{i<j}\frac{1}{|{\mathbf r}_i-{\mathbf r}_j|},
\label{ham}
\end{equation}
where $\omega_0$ defines the strength of the external
confinement, which is assumed to be parabolic in shape. 
The Coulomb potential caused by the 
$N_{\rm imp}$ impurity particles located in the vicinity
of the QD and having charges $|q_k|$, 
is written as
\begin{equation}
V_{\rm imp}({\mathbf r})=\sum^{N_{\rm imp}}_{k=1}\frac{|q_k|}
{\sqrt{({\mathbf r}-{\mathbf R}_k)+d_k}},
\end{equation}
where ${\mathbf R}_k$, and $d_k$ define the lateral and
vertical positions of the particles.
This model has been shown to result in a good agreement with 
the measured transport spectrum of a vertical QD.~\cite{esa6}

In the first calculations, we include a single impurity with a 
negative unit charge on the QD plane ($d=0$) and vary only the lateral
position $R$. Further, when performing a statistical analysis,
we assume that there exist $1-10$ randomly distributed impurity 
particles migrated in the tunneling barrier of the QD. 
We choose the ranges for the impurity positions as 
$0\leq R_k\leq 10\,a^*_B$ and $0\leq d_k\leq 1\,a^*_B$. We set $q_k=-1$ which
is in a qualitative accordance with the results from the tunneling 
experiments.~\cite{esa6} In addition, the minimum
spacing between two impurities is expected to be $0.1\,a^*_B$ which is
of the order of the lattice constant for the layer material 
(e.g., AlGaAs).

\section{Method} \label{method}

The numerical procedure for solving the electronic properties
of the system defined above is based on the spin-density-functional
theory (SDFT). In the self-consistent Kohn-Sham formulation, the
effective Schr\"odinger equation to be solved iteratively is 
written as
\begin{equation}
\left[-\frac{\nabla^2}{2}+V_{\rm KS}({\mathbf r})\right]
\Phi_i^\sigma(\mathbf{r})=\epsilon_i\Phi_i^\sigma(\mathbf{r}),
\label{eq:kse}
\end{equation}
where the Kohn-Sham potential $V_{\rm KS}$ is a sum of the external
potential defined above, the Hartree potential, and the 
exchange-correlation potential given as 
$V_{\rm xc}({\mathbf r})=\delta{E_{xc}}[\rho,\xi]/\delta\rho^\sigma{({\mathbf r})}$. 
Here $\rho$ is the electron density, $\sigma$ denotes the spin index, and
$\xi(\mathbf{r})=[\rho^\uparrow(\mathbf{r})-\rho^\downarrow(\mathbf{r})]/\rho(\mathbf{r})$
is the spin polarization. For $E_{xc}$ we use the local spin-density
approximation based on the functional provided by 
Attaccalite and co-workers.~\cite{attaccalite} 

A detailed description of the following algorithms can
be found in Refs.~\onlinecite{newton,collective,michael}.
We start with solving for the lowest $n$ solutions of the eigenvalue 
problem (\ref{eq:kse}) by applying the evolution operator,
\begin{equation}
\mathcal{T}(\epsilon)\equiv\text{e}^{-\epsilon H}
\end{equation}
repeatedly to a set of states $\{\psi_j,\, 1\leq j\leq n\}$, and 
orthogonalizing the states after every step. Instead of the commonly 
used second-order factorization in combination with the Gram-Schmidt 
orthogonalization, we use the fourth-order factorization for the 
evolution operator~\cite{imstep} given by
\begin{equation}
\mathcal{T}^{(4)}\equiv\text{e}^{-\frac{1}{6}\epsilon V}\text{e}^{-\frac{1}{2}\epsilon T} \text{e}^{-\frac{2}{3}\epsilon\tilde{V}}\text{e}^{-\frac{1}{2}\epsilon T}
\text{e}^{-\frac{1}{6}\epsilon V}=
\text{e}^{-\epsilon[H+\mathcal{O}(\epsilon^4)]}.
\end{equation}
For the orthonormalization, we diagonalize the matrix of overlap 
integrals and construct from these a new set of orthonormal states. 
Depending on the physical system this method can be faster by up 
to a factor of 100 in comparison to the second-order factorization.

A widespread problem of the density-functional calculations 
is the large number of iterations required to obtain
a self-consistent solution. Namely, to maintain the stability of 
the iteration process, it is necessary to keep the mixing 
parameter(s) small, which can lead to thousands of iterations. 
We overcome this problem with a procedure having its roots in 
the Hohenberg-Kohn theorem which states
that the one-body density $\rho(\mathbf{r})$ is the only 
truly independent variable. Therefore an algorithm which solves 
for $\rho(\mathbf{r})$ directly by applying a Newton-Raphson procedure
is used. We define
\begin{equation}
\Delta\rho^\sigma(\mathbf{r})=\sum_\mathbf{h}n^\sigma(\mathbf{h}) 
|\Phi^{\sigma}_{\mathbf{h}}[\rho^\uparrow,\rho^\downarrow](\mathbf{r})|^2
-\rho^\sigma(\mathbf{r})
\end{equation}
as the density difference between two self-consistent iterations.
Here $n^\sigma$ is the occupation factor,
$\Phi^{\sigma}_{\mathbf{h}}$ are the orthogonalized solutions of 
Eq. (\ref{eq:kse}), and $\rho^\sigma(\mathbf{r})$ is the density used 
for the calculation of $V_{\rm KS}$. The sum is over all occupied 
(hole: $\mathbf{h}$) states. Then the density correction 
$\delta\rho^{\sigma}(\mathbf{r})$ is determined by a linear equation
\begin{equation} \label{eq:fullresp}
\Delta\rho^\sigma(\mathbf{r}) = \sum_{\sigma^\prime}\int
d^dr^\prime\,\varepsilon^{\sigma,\sigma^\prime}
(\mathbf{r},\mathbf{r^\prime};0)
\delta\rho^{\sigma^\prime}(\mathbf{r}^\prime). 
\end{equation}
Here $d$ is the dimension of the system and
$\varepsilon^{\sigma,\sigma^\prime}$ is the static 
dielectric function of a non-uniform electron gas.~\cite{PinesNoz}
It contains the zero-frequency Lindhard function and
$V_{p-h}^{\sigma,\sigma^\prime}(\mathbf{r},\mathbf{r^\prime})=
\delta V_{\rm KS}^\sigma(\mathbf{r})/
\delta\rho^{\sigma^\prime}(\mathbf{r^\prime})$.
To avoid the calculation of unoccupied (particle: $\mathbf{p}$) states we seek
for an approximation for the static
response function that only needs the calculation of occupied states.
For that purpose, we recall that linear response theory can be derived
\cite{ThoulessBook} from an action principle for excitations of the
form $\left|\psi(t)\right\rangle = \exp(\sum_{ph}c_{ph}(t)a^\dagger_p
a_h) \left|\phi_0\right\rangle$, where $\left|\phi_0\right\rangle$ is
the ground state, and $c_{ph}(t)$ are particle--hole
amplitudes. If we assume that the particle--hole amplitudes are
matrix elements of a local one-body operator $c^\sigma({\bf r})$, we
end up with Feynman's theory of collective excitations.~\cite{Feynman} 
In this approximation, we can rewrite
Eq.~(\ref{eq:fullresp}) as
\begin{eqnarray}
\left[-\frac{1}{2}\nabla\cdot\left[\rho^\sigma(\mathbf{r})
\nabla\right] +
2\sum_{\sigma^\prime}S_F^\sigma\,\star\,V^{\sigma,\sigma^\prime}_{p-h}
\,\star\,S_F^{\sigma^\prime}\,\star\right] w^{\sigma^\prime} \nonumber \\
= 2\sum_{\sigma^\prime}S_F^\sigma\,\star\, V^{\sigma,\sigma^\prime}_{p-h}
\,\star\,\Delta\rho^{\sigma^\prime},
\label{eq:resp2}
\end{eqnarray}
where now
\begin{equation}
\delta\rho^\sigma(\mathbf{r})\,=\,\Delta\rho^\sigma(\mathbf{r})-
S^\sigma_F(\mathbf{r},\mathbf{r^\prime}) \,\star\,{w}^\sigma(\mathbf{r^\prime}),
\end{equation}
and
\begin{equation}
S_F^\sigma(\mathbf{r},\mathbf{r}^\prime)
=\rho^\sigma(\mathbf{r})\delta(\mathbf{r}-\mathbf{r^\prime})-
\frac{1}{2} \left|\sum_\mathbf{h}\Phi^{\sigma}_{\mathbf{h}}
(\mathbf{r})\Phi^{\sigma}_{\mathbf{h}} (\mathbf{r^\prime})\right|^2,
\end{equation}
which is proportional to the static structure function of the
noninteracting system.
Above, the asterisk stands for the convolution integral. With these
manipulations, we have rewritten the response-iteration equation in a
form that requires only the calculation of the occupied states. 
Since the multiplication of the left-hand side of Eq.~(\ref{eq:resp2}) 
requires only vector--vector operations, the equation can be solved 
with the conjugate--gradient method.

\section{Single impurity} \label{results1}

\begin{figure}
\includegraphics[width=8cm]{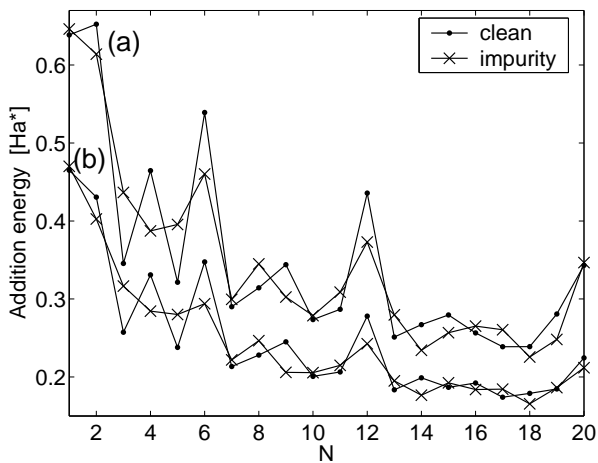}
\caption{Addition energies for clean and impurity-containing QD's
with the confinement strengths of $\omega_0=5$ meV (a) and
$3$ meV (b).}
\label{add1}
\end{figure}

Figure~\ref{add1} shows the addition energies up to 20 electrons
for two QD's with $\omega_0=5$ and 3 meV. A single impurity 
particle with $q=-1$ is located on the QD plane ($d=0$) at $R=2\,l_0$, where
$l_0=\sqrt{1/m^*\omega_0}$, giving $R=3.00\,a^*_B$ and $3.87\,a^*_B$ for
$\omega_0=5$ and 3 meV, respectively. 
The addition energies of the corresponding
clean (no impurity) QD's are also presented. They qualitatively agree 
with the previous SDFT results by Reimann {\em et al.}~\cite{reimann}
and Hirose and Wingreen~\cite{hirose2} for 
$\omega_0\sim 3$ meV [Fig.~\ref{add1}(b)]. Increasing the confinement
strength to $5$ meV leads to the disappearance of the
even peaks at $N=14\ldots 18$, but the ground-state spins remain the same, 
having the well-known sequence determined by Hund's rule.~\cite{hirose2}

In the presence of the impurity, the peaks at magic electron
numbers ($N=2,6,12,20$) are preserved. In contrast, the peaks
at $N=4$ and $9$ which in the clean case correspond to aligned spins 
in the 2nd and 3rd shells, are strongly diminished and
the ground-state spins are $S=0$ and $1/2$, respectively.
Generally, the inclusion of a single impurity
worsens the agreement between the calculated addition-energy spectrum 
and the experimental result by Tarucha and co-workers.~\cite{tarucha}

\begin{figure}
\includegraphics[width=8cm]{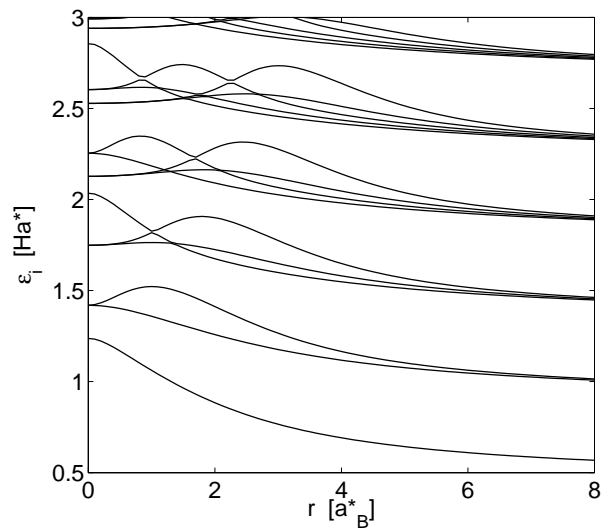}
\caption{Noninteracting single-electron energies for a quantum
dot containing a single repulsive 
impurity particle on the dot plane ($d=0$) 
and at $R=0\ldots 8\, a^*_B$ from the center.}
\label{single}
\end{figure}

Next we study how the energetics depends on the location of the
impurity particle on the QD plane. The results presented below
have been calculated using the fixed confinement strength of 
$\omega_0=5$ meV. Figure~\ref{single} shows the 
noninteracting single-electron energies as a function of $R$.
At large $R$, the energies approach the shells of a clean QD with
$\epsilon_{nl}=(2n+|l|+1)\omega_0$, whereas $R=0$ corresponds
to the eigenenergies of a finite-width quantum 
ring~\cite{chakraborty} with several doubly-degenerated shells.
The mid-$R$ range is characterized by many avoided crossings
between the states that a degenerate in the absence of the impurity.

\begin{figure}
\includegraphics[width=8.5cm]{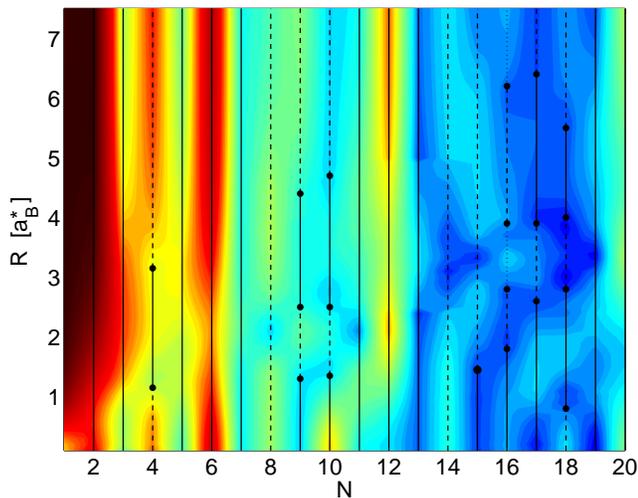}
\caption{(color online) Addition energies (scale from dark blue 
to dark red) for a quantum dot containing a single repulsive 
impurity particle at $R=0\ldots 7.5\, a^*_B$ from the dot center.
The filled circles denote the changes in the ground-state spin.
The solid, dashed, and dotted lines correspond to $S=0$ or $1/2$, $S=1$
or $3/2$, and $S=2$ or $5/2$, respectively.}
\label{add_phase}
\end{figure}

We performed the SDFT calculations for $N=2\ldots 21$ and 
$R=0\ldots 7.5\,a^*_B$
with a step size of $0.1\,a^*_B$. For each $N$ and $R$, we 
calculated several states to find the correct ground-state energy 
and spin. The plot of the addition energies is shown in 
Fig.~\ref{add_phase}.
The values indicated by different colors in the figure vary between 
$0.22\,{\rm Ha}^*$ (dark blue) and $0.65\,{\rm Ha}^*$ (dark red).
The dots with magic electron numbers 
remain particularly stable and paramagnetic ($S=0$) 
despite the impurity particle located at different radius from the center. 
The most apparent exception is $N=12$ at small $R$, showing a shift 
to $N=10$ as a stable configuration. This corresponds to a state with 
filled shells and $S=0$ in the quantum-ring-like geometry, as 
seen in Fig.~\ref{single}. QD's with $N=4,8,14,$ and $18$ have partially
occupied highest shells in the both limits, leading to $S=1$ due to 
Hund's rule. The constant $S=1$ ground state of $N=8$ and $14$ can be
understood from the small energy gap between 
the 4th and 5th, as well as between the 7th and 8th eigenstates 
throughout the $R$ regime (Fig.~\ref{single}).

\begin{figure}
\vspace{0.5cm}
\includegraphics[width=8.5cm]{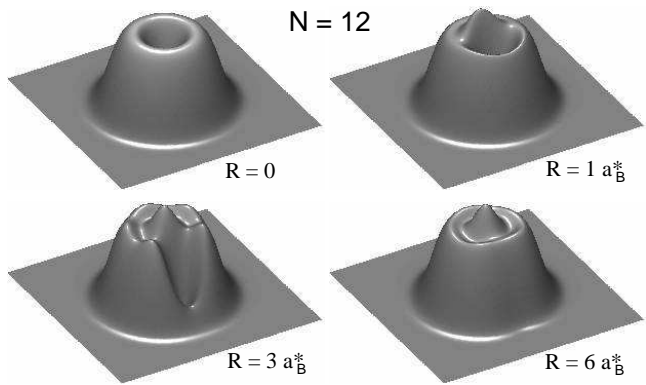}
\caption{Electron densities of a 12-electron quantum dot with
an impurity particle located at $R=0$, $1$, $3$, and $6\,a^*_B$ from the
dot center, respectively.}
\label{dens}
\vspace{0.5cm}
\end{figure}

In Fig.~\ref{dens} we show examples of the 
electron densities for a 12-electron QD
with different impurity locations. The insertion of the impurity particle
at the center ($R=0$) leads to a typical quantum-ring-like solution
with the density distribution on the edge of the QD. As the impurity
is located off the center, the density deforms gradually through
a slightly localized shape with six peaks to a clean
configuration where the circular symmetry is retained. 

We also find many solutions, always corresponding to half-filled 
shells, where the spin symmetry is broken.
These so-called spin-density waves (SDW)~\cite{koskinen}
have been criticized by Hirose and Wingreen~\cite{hirose2}, arguing
that the solutions are mixtures of different (total) spin states,
and thus artifacts of the SDFT. 
Recently, the claim was shown true explicitly in a rectangular 
four-electron QD by employing the SDFT and
the exact diagonalization.~\cite{ari} Nevertheless, the total energy of the 
SDW solution is only slightly smaller than the exact energy
and in any case it represents an improvement over the 
corresponding single-configurational DFT (spin-compensated) energy. 
Hence, the SDW solutions are included as such in the energetics
considered in this paper.

\section{Random impurities} \label{results2}

Then we include a random number (1-10) 
of negatively charged impurities randomly in the 
expected tunneling barrier below the QD (see Sec.~\ref{model} for
the parameters). We computed 1000 impurity configurations such that
for each configuration the number of electrons in the QD was $N=1\ldots 21$.
To determine the groud-state spins in the many-electron problem, we calculated 
all the relevant ($S\leq 2$) spin states for each $N$.

\subsection{Single-electron picture} \label{singlepic}

\begin{figure}
\includegraphics[width=6cm]{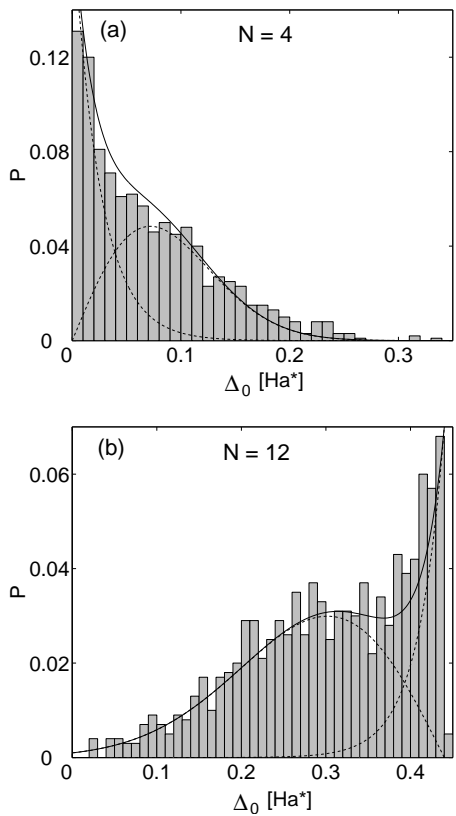}
\caption{Noninteracting addition-energy distributions of 1000
impurity configurations corresponding to 
$N=4$ (a) 
and $N=12$ (b). 
The approximative contributions of the Poisson and Wigner-type 
distributions are also shown. 
The energy binwidth is $0.01\,{\rm Ha}^*$.}
\label{nondis}
\end{figure}

We begin with the level statistics of a QD with {\em noninteracting}
electrons. Now the electrons are affected only by the 
confining and impurity potentials. This corresponds to the
single-electron picture where the addition energy is simply 
equal to the level spacing, $\Delta_0(N)=\epsilon_{N/2+1}-\epsilon_{N/2}$, 
where the divisor of two follows from the spin degeneracy.
Fig.~\ref{nondis}
shows examples of the (noninteracting) addition-energy distributions  
as $N=4$ and 12. 
As the average impurity density is
relatively low, there are several configurations that are close to the
clean case, leading to a Poisson-like level distribution. On the other
hand, the strongly distorted configurations lead to Wigner distribution.
Thus, the sum of these subgroups give the observed double-peak shape
which is pronounced in the 12-electron case [Fig.~\ref{nondis}(b)]. 
The active region of the QD is then larger than that for $N=4$, 
leading to a higher number of distorted configurations since the 
impurity density is kept constant. In both cases, we have divided
the configurations into these two parts in such a way 
that QD's with a total energy higher than a certain, qualitatively 
chosen critical value are considered as ``distorted'' and the 
remaining configurations as ``clean''. The fits corresponding to 
the sums of the Poisson and Wigner functions and $\Delta_0$
is plotted in Fig.~\ref{nondis}. The distributions are mirror images of each
other, since in the clean QD $N=12$ is a filled-shell case 
with $\Delta_0(12)\sim 0.45\,{\rm Ha}^*$, whereas $\Delta_0(4)=0$.

For QD's, the level-spacing distributions are typically
plotted by keeping the QD parameters as the external confinement or
the magnetic field constant and considering high number of eigenstates
within a certain energy range. The integrability of the system can then 
be connected to the level spacings such that the Wigner
distribution is interpreted as a signature of quantum chaos.~\cite{kaaos}
Our case is qualitatively similar in the sense that the varied parameter
is the impurity configuration instead of the energy, and high distortion
leads to the Wigner distribution as shown by Hirose and 
co-workers.~\cite{hirose1}
We suggest, however, that in isolated 
few-electron QD's the average impurity density is 
considerably smaller than in their model, leading to a significant 
contribution of the Poisson-type level-spacing distribution as
visualized in Fig.~\ref{nondis}. We note here that 
Hirose {\em et al.}~\cite{hirose1} defined the noninteracting
picture such that the electrons do not interact with the impurities, while
we consider the eigenenergies of a single electron in the presence of 
the impurity potential.

\subsection{Interacting electrons}

\begin{figure}
\includegraphics[width=8.5cm]{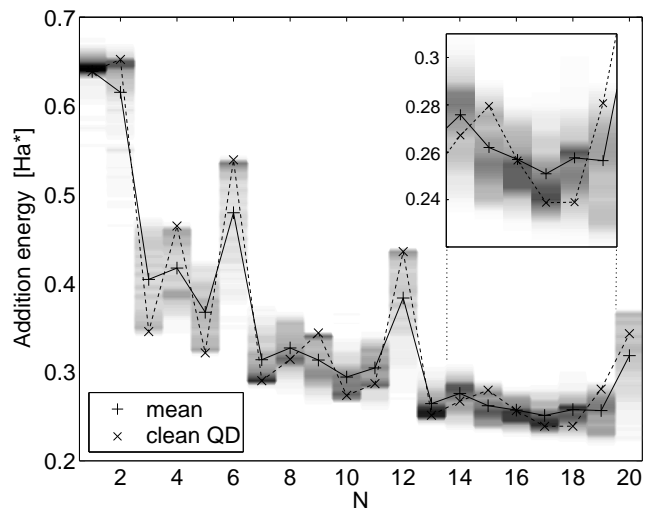}
\caption{Addition-energy distributions of 1000 different
impurity configurations in a quantum dot of 1-20 electrons. 
All the possibilities for the ground-state spins are taken
into account.}
\label{add_dis}
\end{figure}

We examine the corresponding many-electron problem using 
the same 1000 randomly determined impurity configurations. Taking
the different possibilities for each ground-state spin into account,
we performed $\sim 70000$ self-consistent SDFT calculations.
Figure~\ref{add_dis}
shows the addition energy spectrum, where the distribution
for each $N$ is plotted in grayscale. The mean values of the
distributions are shown as pluses, and the crosses correspond
to the clean, parabolic QD [repeated from Fig.~\ref{add1}(a)].
Particularly with small $N$, the result of the clean case matches 
rather accurately with the maxima of the distributions due
to the relatively high fraction of non-distorted
configurations. In the regime $13\leq N\leq 19$, there are clear 
deviations between the clean QD and the maxima, as
visualized in the inset of Fig.~\ref{add_dis} in more detail.
The reason is the high degeneracy of different states in this
regime and the increased tendency to distortion due to the
relatively large active-dot region.

\begin{figure}
\includegraphics[width=6cm]{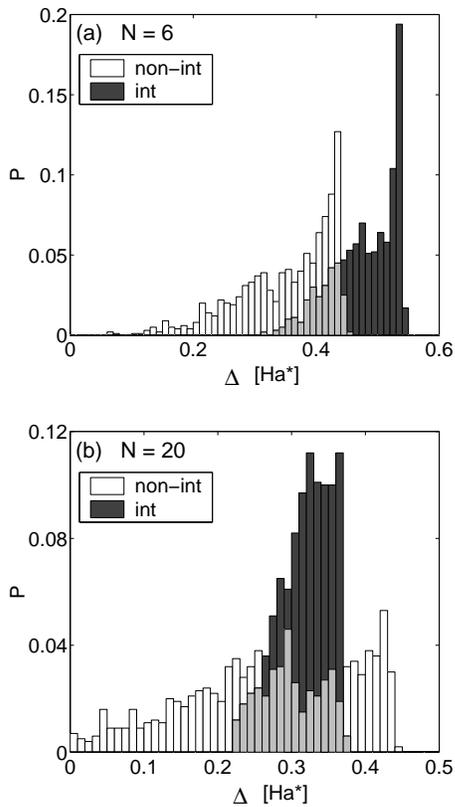}
\caption{Addition-energy distributions for $N=6$ (a) and 
$N=20$ (b) quantum dots in noninteracting (white) and
interacting (dark gray) cases.}
\label{intdis}
\end{figure}

In Fig.~\ref{intdis}
we have a detailed look on the addition-energy distributions
of $N=6$ and $20$ QD's in comparison with the corresponding 
values for noninteracting QD's defined in Sec.~\ref{singlepic}.
We have chosen these magic electron numbers to exclude the
spin effects. Namely, in determining $\Delta(6)$ and $\Delta(20)$,
practically all impurity configurations give $S=0$ and $1/2$ for these 
$N$ and $N\pm 1$, respectively. 
As expected, the Coulomb interaction shifts the 
distributions to higher energies, particularly in the $N=6$
QD where the interaction strength is relatively larger. 
In general, the interactions enhance the average addition energy 
of filled-shell QD's by a factor of $1.17-1.50$
(see Table~\ref{table}).
\begin{table}
\caption{Average noninteracting addition energies $\left<\Delta_0\right>$, their rms 
fluctuations $\delta{\Delta_0}$, the corresponding interacting case with $\left<\Delta\right>$ and $\delta{\Delta}$, 
and the variation coefficient $\delta{\Delta}/\left<\Delta\right>$ for even $N$. The fractions of the
ground-state spins are also shown, excluding $N=16$ quantum dot with $P_{S=2}=0.28$.}
\begin{tabular*}{\columnwidth}{@{\extracolsep{\fill}}|c||c|c|c|c|c|c|c|}
\hline
$N$ & $\left<\Delta_0\right>$ & $\delta{\Delta_0}$ & $\left<\Delta\right>$ & $\delta{\Delta}$ & $\delta{\Delta}/\left<\Delta\right>$ & $P_{S=0}$ & $P_{S=1}$ \\
\hline
2  & 0.411 & 0.0472 & 0.615 & 0.0521 & 0.085 & 1 & 0 \\
4  & 0.070 & 0.0595 & 0.417 & 0.0324 & 0.078 & 0.29 & 0.71 \\
6  & 0.353 & 0.0777 & 0.479 & 0.0511 & 0.107 & 1 & 0 \\
8  & 0.048 & 0.0400 & 0.327 & 0.0193 & 0.059 & 0.11 & 0.89 \\
10 & 0.071 & 0.0552 & 0.294 & 0.0191 & 0.065 & 0.37 & 0.63 \\
12 & 0.306 & 0.0993 & 0.383 & 0.0428 & 0.112 & 1 & 0 \\
14 & 0.042 & 0.0339 & 0.276 & 0.0144 & 0.052 & 0.06 & 0.94 \\
16 & 0.045 & 0.0348 & 0.257 & 0.0113 & 0.044 & 0.24 & 0.48 \\
18 & 0.070 & 0.0517 & 0.258 & 0.0189 & 0.073 & 0.39 & 0.61 \\
20 & 0.271 & 0.1119 & 0.318 & 0.0377 & 0.118 & 0.99 & 0.01 \\
\hline
\end{tabular*}\label{table}
\end{table}
In the mid-shell regime where the level-spacings fluctuate 
from degeneracy, the average noninteracting addition energies 
$\left<\Delta_0\right>$ are considerably lower, i.e., 
$\left<\Delta\right>\sim 3\ldots 7\,\left<\Delta_0\right>$.

As seen in Fig.~\ref{intdis} and Table~\ref{table},
interactions decrease the fluctuation of the 
addition energies defined as 
$\delta{\Delta}=\sqrt{\left<(\Delta-\left<\Delta\right>)^2\right>}$.
The result is in contrast with strongly disordered
QD's~\cite{hirose1} and can be explained 
with the shell structure determined
by the primarily symmetric (parabolic) confining potential:
in the interacting system the impurities are screened 
by the electrons such that the QD is effectively more
symmetric than in the corresponding noninteracting case,
which leads to enhanced stability and smaller fluctuations.
The variation coefficients 
$\delta{\Delta}/\left<\Delta\right>$ show that the 
fluctuations are relatively largest for the magic electron 
numbers, since the configurations with
high disorder have the strongest distorting effect on 
the corresponding filled-shell structure.



\begin{figure}
\includegraphics[width=8.5cm]{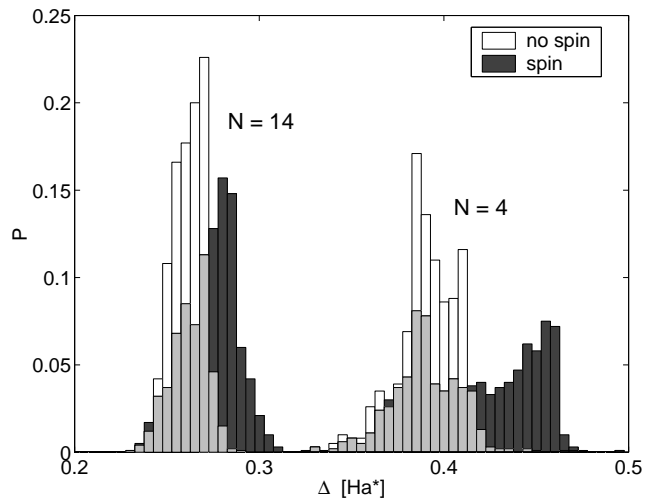}
\caption{Addition-energy distributions for two quantum dots of 
4 and 14 interacting electrons, respectively. 
The results have beed calculated both by (i) taking 
all the possible ground-state spins into account (dark gray)
and (ii) neglecting Hund's rule and thus setting $S=0$ and $1/2$
for the ground states of $N=4,14$ and $N=3,5,13,15$, respectively (white).}
\label{spindis}
\end{figure}

Figure~\ref{spindis}
shows the addition-energy distributions for $N=4$ and $14$ QD's
when the different ground-state spins are included (dark grey) and
neglected (white). With both $N$, there is (approximately) 
an equal possibility for $S=1$ ($\sim$ clean) and $S=0$ (distorted),
leading to a larger width of the distribution. When $N=4$, the
splitting of the different spin states into two separate
peaks is clear. Hence, the measured addition energy may depend strongly 
on the impurity configuration of the particular sample.

Fractions of $S=0$ and $S=1$ ground states in the calculated 
configurations for even $N$ are shown in Table~\ref{table}.
Except a few results in $N=12$ QD and $\sim 1\,\%$ of
the configurations in $N=20$ QD, the magic electron numbers
correspond to $S=0$ states. 
Otherwise, the $S=1$ states formed by Hund's rule are dominant.
The $N=16$ QD, which in the clean case 
has a half-filled shell with a four-fold degeneracy, has
also a significant fraction of $S=2$ states.

We underline that due to the primarily parabolic confining
potential and a relatively small number of impurities,
the shell structure is generally a dominating feature in the
energetics. However, the features of disorder are clearly
visible in the results: (i) A considerable fraction of the 
configurations shows Wigner shape in the level-spacing
distribution (Fig.~\ref{nondis}); (ii) the relative fluctuation in 
$\left<\Delta\right>$ is largest for filled-shell QD's; (iii)
interactions bring Gaussian contribution to the distribution
functions in the mid-shell regime (Fig.~\ref{spindis}).
The last feature is typical for chaotic and disordered QD's
with considerable electron-electron interactions.~\cite{berkovits}

\section{Summary} \label{summary}

To summarize, we have used a fast response-function algorithm within
the spin-density-functional theory to study the stability of the shell
structure in few-electron quantum dots. A single impurity near the 
quantum dot may cause dramatic changes in the addition energies, but
the shell structure is generally preserved with high peaks at
the magic electron numbers.
The same effect was found using a model of one thousand random
impurity configurations, even if the fluctuations are largest in the 
filled-shell regime. In the noninteracting picture, the 
addition-energy distributions are combinations of Poisson and Wigner 
forms. The electron-electron interactions induce a shift to higher
energies and considerably smaller fluctuations.
In the mid-shell regime, changes in the ground-state spin lead to a strong 
broadening of the distribution. Hence, in a series of measurements there can be
large fluctuations in those addition energies, and the result for
a particular quantum dot strongly depends on the impurities 
in the sample.

\begin{acknowledgments}
We thank E. Krotscheck, M. J. Puska, and A. Harju for fruitful discussions.
This work was supported by the Austrian Science Fund under project P15083-N08
and by the Academy of Finland through its Centers of Excellence 
program (2000-2005).
\end{acknowledgments}


\begin{thebibliography}{10}
\bibitem{units}
With our material parameters, the effective Hartree ${\rm Ha}^*\approx 11.30$
  meV and the effective Bohr radius $a^*_B\approx 10.03$ nm.
\end{thebibliography}

\end{document}